\documentclass[aps, prl, twocolumn, preprintnumbers, amsmath, amssymb, superscriptaddress, longbibliography]{revtex4-2}

\usepackage[colorlinks=true,allcolors=blue]{hyperref}
\usepackage{txfonts}
\usepackage{graphicx}
\usepackage{color}
\usepackage{dcolumn}
\usepackage{bm}
\usepackage{feynmp}
\usepackage{braket}
\usepackage{ulem}

\begin{document}

\title{Spin Wave Propagation through Antiferromagnet/Ferromagnet Interface}
\author{Oksana Busel}
\email{Oksana.Busel@oulu.fi}
\affiliation{Nano and Molecular Systems Research Unit, University of Oulu, Oulu 90014, Finland}
\affiliation{Faculty of Physics and Mathematics, National Technical University
of Ukraine ``Igor Sikorsky Kyiv Polytechnic Institute,'' Kyiv 03056, Ukraine}
\author{Oksana Gorobets}
\affiliation{Faculty of Physics and Mathematics, National Technical University
of Ukraine ``Igor Sikorsky Kyiv Polytechnic Institute,'' Kyiv 03056, Ukraine}
\author{Oleg~A.~Tretiakov}
\email{o.tretiakov@unsw.edu.au}
\affiliation{School of Physics, The University of New South Wales, Sydney 2052, Australia}

\date{December 29, 2021}

\begin{abstract}
We study the problem of controlling spin waves propagation through an antiferromagnet/ferromagnet interface via tuning material parameters. It is done by introducing the degree of sublattice noncompensation of antiferromagnet (DSNA), which is a physical characteristic of finite-thickness interfaces. The DSNA value can be varied by designing interfaces with a particular disorder or curvilinear geometry. We describe a spin-wave propagation through any designed antiferromagnet/ferromagnet interface considering a variable DSNA and appropriate boundary conditions. As a result, we calculate the physical transmittance and reflectance of the spin waves as a function of frequency and show how to control them via the exchange parameters tuning.
\end{abstract}

\maketitle

Antiferromagnets (AFMs) are attractive as active elements for next-generation spintronic devices \cite{MacDonald2011,Chumak2015,Baltz2018,Ni2021}. The disadvantage of the last generation of spin-wave (SW) devices is their relatively low operating frequency, which is limited by the GHz range \cite{Mohseni2021,Yu2016,Sampo2018,Liu2018,Kim2020}. However, recent trends involving AFMs allow to raise this frequency to THz  to successfully compete with optical devices \cite{Gobel2021,Mashkovich2019,Dabrowski2020,Cenker2021}, because a spin manipulation in AFMs is inherently faster than in ferromagnets (FM) \cite{rezende2020,nemec2018}. The ultrafast generation, detection, and gate tuning of spin waves in AFMs are now becoming more and more accessible \cite{nemec2018,zhang2020}, meanwhile the detection techniques in FMs are still better established \cite{maendl2017,gubbiotti2010,ulrichs2018}.

The benefits of using AFMs have been recently actively studied~\cite{yin2019,zelezny2014,Cheng2014,Shick2010,Wadley2017,Shen2020,wang2019,Scott1985,DuttaGupta2020}. Usually, AFMs are considered in limiting cases, namely, compensated, when the AFM has no static magnetization at the interface \cite{Mikhailov2017}, and noncompensated, when the boundary of the AFM is magnetized \cite{Gruyters2008}. AFMs with the fully compensated spin moments have implementations in terahertz range \cite{puliafito2019,kampfrath2010,tveten2016}, e.g., spin-current driven AFM nano-oscillators have been proposed \cite{lee2019,Shen2019}. On the other hand, the mechanisms of the AFM domain switching by current pulses \cite{chen2018,shiino2016,baldrati2019} in certain cases are based on the electric field or spin torques acting on the noncompensated spins at the AFM interface \cite{Belashchenko2016,moriyama2018}. However, in some problems such as transmission of the spin current through a thin AFM insulator by means of evanescent AFM SWs, it has been recently assumed that the AFM could be partially noncompensated at the interface \cite{khymyn2016}. Moreover, it has been pointed out that the effective magnetic moment can be a complex arrangement of not fully compensated moments along compensated AFM structures \cite{buchner2019}. Considering the compensated or noncompensated limiting cases significantly simplifies the description of underlying AFM physics, but does not cover the entire spectrum of the AFMs, and furthermore, requires perfectly flat AFM interfaces. Meanwhile, realistic rough interfaces with a variable degree of sublattice noncompensation are more common and therefore important to study.

In this paper, we investigate the SWs propagation through AFM/FM interface of a complex geometry. This problem can be solved by introducing an averaged characteristic of this interface -- degree of sublattice noncompensation of AFM (DSNA). Due to strong exchange interactions between the  AFM sublattices and with a neighboring material, this concept may play a crucial role in describing complex AFM interfaces with other magnetic media, i.e., those with a specific disorder or of curvilinear geometry. Complex interfaces with variable DSNA can be manufactured in a variety of ways, such as by engineering a particular surface roughness, ion implantation, vapor deposition or introduction of other defects \cite{Tretiakov2016,Xue2015,Li2013,Slezak2019,Demeter2012}. 

\begin{figure}[tbp]
\centering
\includegraphics[width = 1\linewidth]{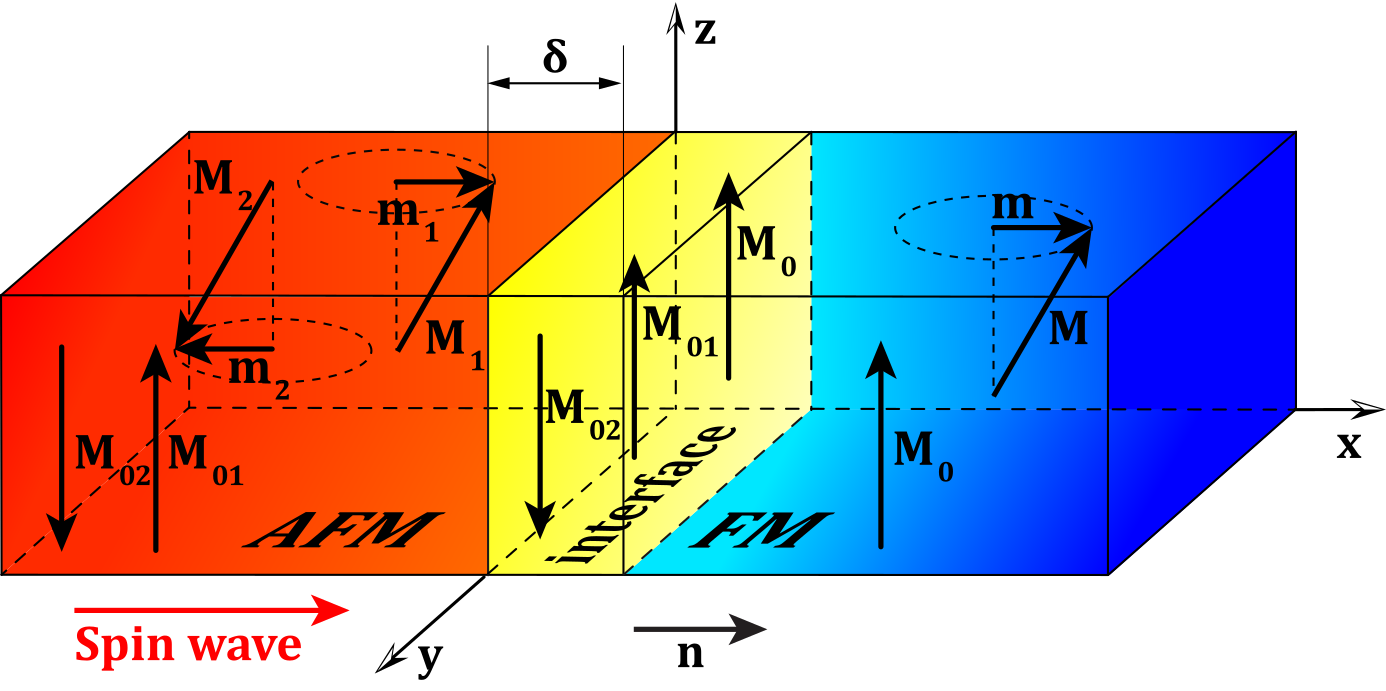}
\caption{A schematic representation of two-sublattice AFM, the interface of finite thickness between AFM and FM, FM and magnetizations in each layer.}
\label{fig:fig1}
\end{figure}

We develop an analytical approach to a SW propagation from AFM to FM through any designed AFM/FM interface described by a variable DSNA.  We consider the scattering of exchange SWs at the AFM/FM interface of finite thickness sandwiched between AFM and FM. We obtain the complete relations between the phases and amplitudes of scattered SWs taking into account the Poynting vector continuity at the boundary. This theory is designed for an exchange dominated regime, in which we derive the interface boundary conditions in the vicinity of noncompensated case, i.e., the DSNA $\mu\to 0$. This case is of high interest as it gives rise to the largest exchange bias in AFM/FM structures.As a result, the frequency dependences of SW transmittance and reflectance at the interface are found in $\mu\to 0$ case. 

\textit{Model.} We consider a one-dimensional scenario of SW propagation through an AFM/FM interface (along the $x$-axis) as shown in Fig.~\ref{fig:fig1}. The interface has thickness $\delta$ and is parallel to the $y$-$z$ plane. Since AFM has two sublattices and FM has one, we introduce three static (ground state) magnetizations within the interface and its surroundings, $\mathbf{M}_{01}$, $\mathbf{M}_{02}$, and $\mathbf{M}_{0}$, respectively. The media are magnetized by a strong enough uniform external magnetic field along the $z$-axis, so all ground state magnetizations are (anti)parallel to the $z$-axis. The small deviations of the magnetizations $\mathbf{M}_{1(2)}$ and $\mathbf{M}$ from the ground state are introduced in the form $\mathbf{M}_{1(2)}=\mathbf{M}_{01(02)}+\mathbf{m}_{1(2)}$ and $\mathbf{M}=\mathbf{M}_{0}+\mathbf{m}$, where $\mathbf{m}_{1(2)}$ and $\mathbf{m}$ are the dynamical components of the magnetization for the two AFM sublattices and FM, respectively, i.e., $m_{1(2)}\ll M_{01(02)}$ and $m\ll M_{0}$.

To present the analytical theory of the SW propagation through the AFM/FM interface in an exchange dominated regime, only the exchange interactions between AFM and FM have to be taken into account within the interface \cite{kruglyak2014,cochran1992,klos2018}. We describe the magnetization dynamics by the Landau-Lifshitz (LL) equation \footnote{The dissipation generally decreases the amplitudes of SWs, but does not influence their behavior qualitatively. Therefore, in the following we can use the LL equation approximation, where the dissipation is neglected.}, $\partial\mathbf{M}_{l}/\partial t=\gamma\mathbf{M}_{l}\times\mathbf{H}^{\rm{eff}}_{l}$, where $\gamma$ is the gyromagnetic ratio and $\mathbf{H}^{\rm{eff}}_{l}=-\delta W_{i}/\delta\mathbf{M}_{l}$ is the effective magnetic field in each material. Here $W_{i}$ is the interfacial magnetic energy, and labels $l$ represent the two AFM sublattices and the one of FM. Integrating over the interface one finds the energy $W_{i}=S\int_{0}^{\delta}w_{i}dx$ with $S$ being the cross-sectional area and the energy density at the interface $w_{i}$:
\begin{eqnarray}
w_{i} & = & A(x) \mathbf{M}_{1} \mathbf{M}_{2} + A_{1}(x)\mathbf{M}_{1} \mathbf{M} + A_{2}(x) \mathbf{M}_{2}\mathbf{M} +\frac{\alpha(x)}{2}\left(\frac{\partial\mathbf{M}}{\partial x}\right)^{2}\nonumber\\
&&+\frac{\alpha_{1}(x)}{2}\left[\left(\frac{\partial\mathbf{M}_{1}}{\partial x}\right)^{2}+\left(\frac{\partial\mathbf{M}_{2}}{\partial x}\right)^{2}\right] 
+{\alpha_{2}(x)}\frac{\partial\mathbf{M}_{1}}{\partial x}\frac{\partial\mathbf{M}_{2}}{\partial x}.
\label{eq:eq1}
\end{eqnarray}
Here $A(x)$ is the homogeneous AFM exchange parameter, $A_{1(2)}(x)$ is the exchange coupling parameter of each AFM sublattice with the FM, $\alpha=\alpha_{\rm{ex}}/M_{0}^{2}$ is the inhomogeneous FM exchange parameter, $\alpha_{1}$ is the inhomogeneous exchange self-interaction parameter for each AFM sublattice, and $\alpha_{2}$ is the inhomogeneous exchange interaction parameter between the AFM sublattices ($\alpha_{n}=\alpha_{{\rm ex}, n}/M_{0n}^{2}$ with $n=1,2$ and $\alpha_{{\rm ex}, n}$ being the exchange stiffness constants \cite{baryakhtar1985}). Parameter $A$ can be estimated by $A\sim\alpha_{{\rm ex}, 1}/(M_{01(02)}^{2}d^{2})$, where $d$ is the lattice constant of the AFM \cite{baryakhtar1980,baryakhtar1985,tveten2013}. The interactions in Eq.~(\ref{eq:eq1}) include the antiferromagnetic coupling between two sublattices, the two couplings between each AFM sublattice and the FM~\cite{puliafito2019}, the exchange stiffness interaction of FM, the exchange self-interaction of the AFM sublattices, and the exchange interaction between AFM sublattices, respectively. 

\begin{figure}[tbp]
	\centering
	\includegraphics[width = 0.98\linewidth]{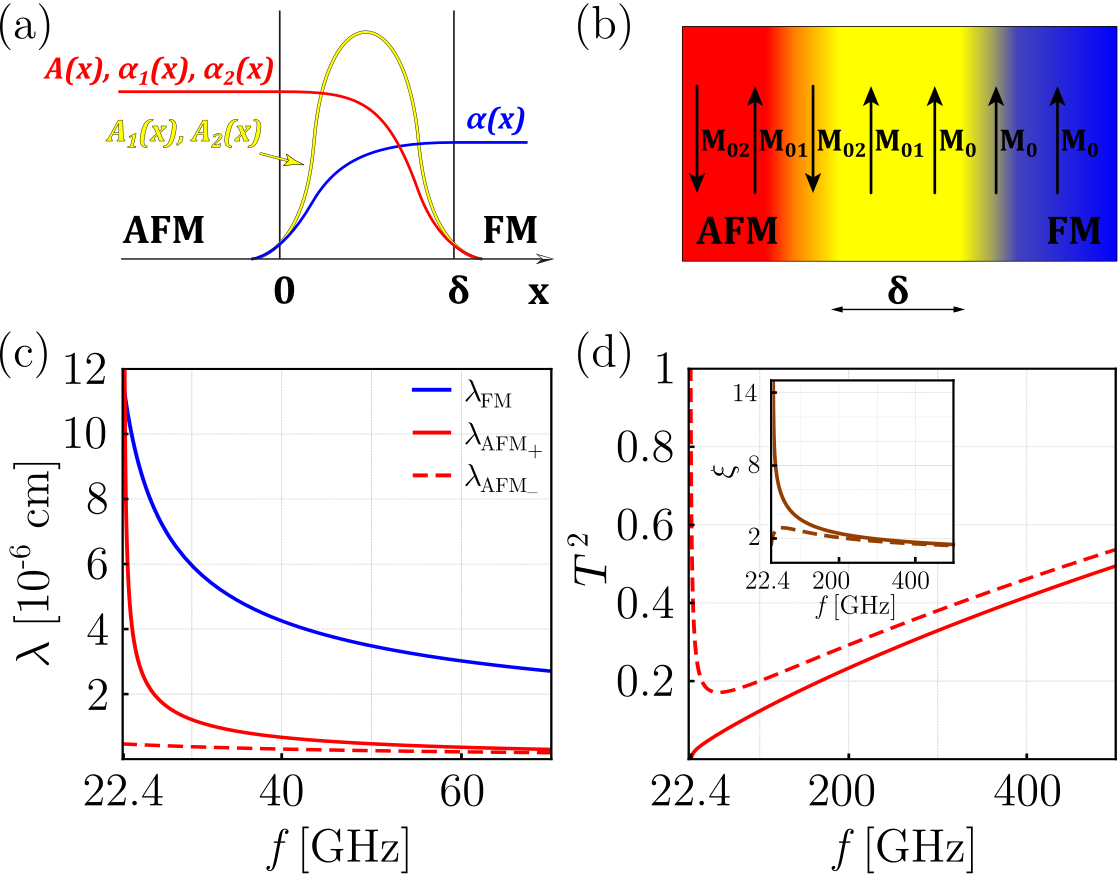}
\caption{(a) Schematic spatial dependence of all material parameters characterizing AFM/FM interface, Eq.~(\ref{eq:eq1}). (b) One-dimensional model for noncompensated AFM at the AFM/FM interface. (c) The dependence of the SW wavelength $\lambda$ and (d) normalized transmittance $T^2$ on frequency $f$. In (c), the solid blue curve represents the FM branch, while in (c) and (d) the solid and dashed red curves correspond to the plus and minus AFM branches, respectively. In (d), $T^2(f)$ is given according to Eq.~(\ref{eq:eq12}) without the influence of the additional factor $\xi$. The inset in (d) illustrates $\xi (f)$, according to Eq.~(\ref{eq:eq17}).}
\label{fig:2_a_b(inset)} 
\end{figure}

\textit{AFM/FM boundary conditions.} To derive the interfacial boundary conditions, order parameters $\mathbf{M}_{1}$, $\mathbf{M}_{2}$ and $\mathbf{M}$ are considered as slowly varying functions, whereas the coefficients $A_{1}(x)$, $A_{2}(x)$, $A(x)$, $\alpha_{1}(x)$, $\alpha_{2}(x)$, and $\alpha (x)$ are varying significantly within the interface $[0,\delta ]$ as shown in Fig.~\ref{fig:2_a_b(inset)}~(a). We define the interface in terms of the averaged properties of the surrounding materials, so the LL equations are integrated over the thickness $\delta$ of the interface \cite{Barnas1991,kruglyak2014,busel2018,Busel2019}. Then at the interface, the solutions of the LL equations satisfy the boundary conditions:
\begin{subequations}\label{eq:eq2}
\begin{align}
\begin{split}
A_{1}m+\left(A-\nu A_{1}-\alpha_{1}\frac{\partial}{\partial x}\right)m_{1}+\left(A-\alpha_{2}\frac{\partial}{\partial x}\right)m_{2}=0
\end{split}, \\
\begin{split}
A_{2}m+\left(A+\nu A_{2}-\alpha_{1}\frac{\partial}{\partial x}\right)m_{2}+\left(A-\alpha_{2}\frac{\partial}{\partial x}\right)m_{1}=0
\end{split}, \\
\begin{split}
\left(A_{1}-A_{2}-\nu\alpha\frac{\partial}{\partial x}\right)m-\nu A_{1}m_{1}-\nu A_{2}m_{2}=0,
\end{split}
\end{align}
\end{subequations}
for the dynamical magnetization components in FM, $\mathbf{m}=(m_{x},m_{y},0)$, and AFM, $\mathbf{m}_{1(2)}=(m_{1(2),x},m_{_{1(2)},y},0)$. For convenience we expressed them via the cyclic variables $m_{1(2)}= m_{1(2),x}+im_{1(2),y}$ and $m=m_{x}+im_{y}$. We also used $M_{02}/M_{01}=1$ for the AFM and introduced the notation $M_{0}/M_{01}=\nu$. Eqs.~(\ref{eq:eq2}) represent the set of linearized boundary conditions for the dynamical magnetization components. Note, that here we have extended the approach of finding boundary conditions for an FM/FM interface \cite{kruglyak2014}. Thus, in our case of the AFM/FM interface, the approach works when the absolute values of the magnetization are approximately equal, i.e. $\nu \approx 1$, and the magnetization of the AFM sublattice closest to the FM boundary is parallel to the FM magnetization, see Fig.~\ref{fig:2_a_b(inset)}~(b), that is called noncompensated AFM boundary case. Since the AFM dynamical magnetization components are strongly coupled, the contribution from both AFM sublattices is effectively included in the interface, but boundary conditions (\ref{eq:eq2}) are greatly simplified.

To obtain the ratio of the AFM dynamical components $\mathbf{m}_{2}/\mathbf{m}_{1}$, we solve the set of linearized LL equations for both sublattices \footnote{see Sec.~I of the Supplemental Material \cite{SupplMat}} and take $M_{01,z}=M$ and $M_{02,z}=-M$, then this ratio takes the form 
\begin{equation}
\sigma(\omega) = -\frac{m_{2}}{m_{1}} = \frac{(\Omega_{+}+\Omega_{-})\omega-\Omega_{+}\Omega_{-}-\omega^{2}}{\Omega_{-}^{2}-\omega^{2}},
\label{eq:eq3}
\end{equation}
where $\Omega_{\pm} =\gamma M[(\alpha_{1}-\alpha_{2})k_{AFM}^{2}\pm H_{e}/M+\beta_{1}-\beta_{2}]$, $H_{e}$ is the external magnetic field, $k_{AFM}$ is the AFM wavevector, $\beta_{1}$ and $\beta_{2}$ are the AFM anisotropy constants, and $\omega=2\pi f$ with frequency $f$. Note that for a nonzero magnetic field (along $z$-axis), the AFM magnetic moments are oriented almost perpendicular to the magnetic field with a slight misalignment along the magnetic field. Thus, in the ground state a small uncompensated magnetic moment would be parallel to the field, while the N\'eel vector would be perpendicular to it. Therefore, to obtain the desired AFM orientation of Fig.~\ref{fig:2_a_b(inset)}~(b), it is known that a strong enough additional easy-axis anisotropy along $z$-axis has to be included.

\textit{DSNA.} Using $\sigma$ from Eq.~(\ref{eq:eq3}), the linearized boundary conditions~(\ref{eq:eq2}) can be simplified \cite{SupplMat}. The set of these equations has a solution with a nonzero amplitude of the transmitted SW only if the following relations are satisfied: $\mu(\alpha_{1}-\alpha_{2} \sigma)=\alpha_{2}-\alpha_{1} \sigma$, $\mu[A(1-\sigma)-A_{1}\nu]=A(1-\sigma)-A_{2} \sigma \nu$, and $\mu A_{1}=A_{2}$ with $\mu\neq0$. Otherwise, the SWs are fully reflected. Thus, we introduce the DSNA $\mu$ as
\begin{equation}
\mu=\frac{\alpha_{2}-\alpha_{1}\sigma}{\alpha_{1}-\alpha_{2}\sigma}.
\label{eq:eq6}
\end{equation}
Then the coupling parameters between each AFM sublattice and FM, $A_{1(2)}$, are defined through the one between the AFM sublattices, $A$, and the DSNA $\mu$ as
\begin{equation}
A_{1}(\mu)=\frac{A(\mu-1)}{\mu \nu},~A_{2}(\mu)=\mu A_{1}(\mu),
\label{eq:eq7}
\end{equation}
when $\sigma\neq 1$. Note that Eq.~(\ref{eq:eq3}) is applicable anywhere in the AFM, whereas Eq.~(\ref{eq:eq6}) is applicable only at the AFM/FM interface. For the noncompensated AFM boundary case we discussed above, Eqs.~(\ref{eq:eq6}) and~(\ref{eq:eq7}) imply that $|A_1| \gg |A|\sim |A_2|$ and $\mu\to 0$.  

To derive the boundary conditions for any designed boundary depending on the DSNA, we should take into account in Eqs.~(\ref{eq:eq2}) the relation between AFM dynamical components [Eq.~(\ref{eq:eq3})] and $A_{1,2}(\mu)$ from Eq.~(\ref{eq:eq7}). This significantly simplifies Eqs.~(\ref{eq:eq2}):
\begin{subequations}\label{eq:eq9}
\begin{align}
\begin{split}
A(1- \mu)m -\nu\left[A(1-\sigma \mu)-\mu({\alpha_1}-{\alpha_2}\sigma )\frac{\partial}{\partial y}\right]{{m}_{1}}=0,
\end{split} \\
\begin{split}
\left[A{{(1-\mu)}^{2}}+{{\nu}^{2}}\mu\alpha\frac{\partial}{\partial y}\right]m -\nu A(1-\mu)(1-\sigma\mu){{m}_{1}}=0.
\end{split}
\end{align}
\end{subequations}
These improved boundary conditions include only one dynamical AFM component, however preserve the information of both AFM sublattices. They are defined for any AFM/FM interface depending on DSNA $\mu$. Thus, when the DSNA is varied, in fact, it changes the ratio of considered spins from the first and second AFM sublattices within the interface.

\textit{SW propagation.} Next, we consider transmission of SWs from AFM to FM. In this paper we assume the geometric optics approximation for the SWs, since the typical SW wavelengths $\lambda_{sw}$ in AFM and FM are much larger than the width of the interface, $\lambda_{sw}\gg \delta\sim d$. We look for a solution where the incident and reflected circularly polarized SWs in the AFM and transmitted SW in the FM are monochromatic plane waves $\mathbf{m}(\mathbf{r},t)=\mathbf{m}(\mathbf{r})\, e^{i\omega t}$ with the dynamical components of the magnetization defined as
\begin{equation}
\label{eq:eq10}
m=\tilde{t}\,e^{ik_{FM,x}x},\!\!\quad
m_{1(2)}=\tilde{I}_{1(2)}\,e^{ik_{AFM,x}x}+\tilde{r}_{1(2)}\,e^{-ik_{AFM,x}x}.
\end{equation}
Here $k_{FM}$ is the FM wavevector, $\tilde{I}_{1(2)}$ is the amplitude of the wave incident onto the first (second) AFM sublattice, $\tilde{r}_{1(2)}=\tilde{R}_{1(2)}e^{i\varphi_{1(2)}}$ and $\tilde{t}=\tilde{T}e^{i\varphi}$ are the complex amplitudes of the reflected wave from first (second) AFM sublattice and transmitted wave into the FM, respectively, where $\tilde{R}_{1(2)}$ and $\tilde{T}$ are the real amplitudes, while $\varphi_{1(2)}$ and $\varphi$ are the phase shifts. Since the problem is formulated in a stationary state, the explicit dependence on time can be neglected here.

Using Eq.~(\ref{eq:eq3}) the relations between the amplitudes and phase shifts of the AFM sublattices take the form $\tilde{I}_{2}=-\sigma\tilde{I}_{1}$, $\tilde{R}_{2}=-\sigma\tilde{R}_{1}$, and $\varphi_{1}=\varphi_{2}$, respectively. In the following it is convenient to introduce dimensionless complex amplitudes $r_{1}=\tilde{r}_{1}/\tilde{I}_1$ and $t=\tilde{t}/\tilde{I}_1$, then using boundary conditions~(\ref{eq:eq9}) and expressing $\sigma(\mu)$ from Eq.~(\ref{eq:eq6}), we obtain $r_1 (\mu)$ and $t (\mu)$ shown in \footnote{See Sec.~II of the Supplemental Material \cite{SupplMat}}. Note that for small $\mu\neq 0$, one can probe different degree of noncompensation by varying $\mu$, while at $\mu\to 0$ we approach a fully noncompensated AFM boundary case, where our approximation works the best~\footnote{We consider the interface as a composite material that includes three magnetizations - two from AFM sublattices and one from FM, as shown in Fig.~(\ref{fig:fig1}). However, according to \cite{kruglyak2014} for finding the boundary conditions in the form of Eqs.~(\ref{eq:eq2}), the only case when magnetization of one of AFM sublattices is parallel and comparable in magnitude to the FM magnetization is appropriate, i.e., noncompensated AFM boundary case $\mu\to 0$.}. In this limit the amplitudes take the form
\begin{eqnarray}
r_{1} &=&-\frac{A\nu^2\alpha\alpha_1 k_{FM}-A(\alpha_1^2-\alpha_2^2)k_{AFM}}{A\nu^2\alpha\alpha_1 k_{FM}+A(\alpha_1^2-\alpha_2^2)k_{AFM}},
\label{eq:eq11}\\
t &=&\frac{2 A\nu(\alpha_1^2-\alpha_2^2)k_{AFM}}{A\nu^2\alpha\alpha_1 k_{FM}+A(\alpha_1^2-\alpha_2^2)k_{AFM}}.
\label{eq:eq12}
\end{eqnarray}
Thus, to obtain dimensionless reflectance $R_{1}^{2}$ and transmittance $T^{2}$ one has to find $|r_{1}|^{2}$ and $|t|^{2}$, respectively. 

To determine the spectrum of SWs in AFM and FM we employ well-known dispersion relations for AFM and FM (we consider the case when AFM and FM have anisotropy of the easy-axis type, i.e., $\beta > 0$, $\beta_{1} -\beta_{2}>0$, and the total magnetization in AFM is small)~\cite{Akhiezer1968},
respectively:
\begin{eqnarray}
\label{eq:eq15}
\!\!\!\!\omega_{\pm}(\mathbf{k}_{AFM})&=&\gamma\left(\sqrt{2ak_{AFM}^{2}+H_{a}^{2}}\pm H_{e}\right),\\
\omega(\mathbf{k}_{FM})&=&\gamma\left(2\alpha M_{0}k_{FM}^{2}+M_{0}\beta+H_{e}\right),
\label{eq:eq16}
\end{eqnarray}
where $a=A(\alpha_{1}-\alpha_{2})M^{2}$, $H_{a}=M\sqrt{2A(\beta_{1}-\beta_{2})}$ is the anisotropy field of AFM, $\beta=K/M_{0}^{2}$ is the anisotropy constant of FM, and $\omega=2\pi f$ with frequency $f$. The AFM has two (plus and minus) dispersion branches, $\omega_{\pm}$. Here we used the dispersion relations for the case of normal incidence of the SW on the AFM/FM interface, the general case of any angle between the SW direction and the interface normal is considered in \cite{SupplMat}. According to Eq.~(\ref{eq:eq15}), to have propagating SWs in the AFM one has to apply a frequency above $\gamma(H_{a} \pm H_{e})/2\pi$. To activate both AFM branches one has to apply a frequency above $f_{a}=\gamma(H_{a}+ H_{e})/2\pi$. 

Assuming that the energy is conserved at the AFM/FM interface~\footnote{In realistic systems, the energy is lost at the boundary, however, generally this loss would not effect qualitatively the results of the proposed model if taken into account.}, the normal component of the energy flux density vector should be continuous at this boundary. In the problem of an interface of two FMs \cite{kruglyak2014}, the number of variables that have to be matched at the boundary is the same on both sides, since the FMs have only one magnetization sublattice. Then equating the normal components of the Poynting vectors of both FMs, the sum of the transmittance $T^2$ and reflectance $R^2$ at the boundary is $T^2+R^2 =1$. However, for an AFM/FM interface, the number of variables to the right and left of the boundary is different, since the AFM has two sublattices (while the FM has only one), see Fig.~\ref{fig:2_a_b(inset)}~(b). In this case, one needs to consider the Poynting vector contributions from both AFM sublattices, which significantly complicates the problem. Thus, for the AFM/FM interface, equating the normal components of the Poynting vector of each medium, introduces the factor $\xi$ responsible for the energy conservation at the interface as~\footnote{For the derivation of $\xi$ see Sec.~IV of the Supplemental Material \cite{SupplMat}.}:
\begin{equation}
\label{eq:eq17}
R_{1}^{2}+\xi T^{2}=1,\quad \xi=\frac{\alpha k_{FM}}{k_{AFM}[\alpha_{1}(1+\sigma^{2})-2\alpha_{2}\sigma]}.
\end{equation}
Noting that at the boundary the ratio $\sigma$ depends on the DSNA $\mu$ according to Eq.~(\ref{eq:eq6}), $\xi (\mu)$ takes the form shown in \footnote{See Sec.~IV of the Supplemental Material~\cite{SupplMat}}, which in the most relevant fully noncompensated case ($\mu=0$) gives $\xi= \alpha \alpha_{1} k_{FM}/[(\alpha_{1}^2-\alpha_{2}^2) k_{AFM}]$.

To understand the transmittance behavior, we consider its dependence on various magnetic parameters. We take the parameters for a typical AFM/FM interface such as NiO/CoFeB: $M=0.84$~kG, $M_{0}=1.2$~kG, $\alpha_{ex,1}=2.8\times10^{-8}$~erg$/$cm, $\alpha_{ex}=2\times10^{-6}$~erg$/$cm, $d=4.2\times10^{-8}$~cm \cite{Tveten2014}, $H_{e}=5$~kOe, $H_{a}=3$~kOe, $K=2.4\times10^{6}$~erg$/$cm$^{3}$. To define the proper frequency range for the SWs propagating in the AFM/FM system we determine the reasonable SW wavelengths of AFM $\lambda_{AFM\pm}=2\pi/k_{AFM}$ and FM $\lambda_{FM}=2\pi/k_{FM}$ by matching the frequencies of the AFM and FM from their dispersion relations, Eqs.~(\ref{eq:eq15}) and~(\ref{eq:eq16}).  For the above parameters, the frequency when the both AFM branches are activated is $f_{a}=22.4$ GHz (below $f_a$ the SW propagation is possible only from the ``minus'' AFM branch). By increasing $f$ to THz range \footnote{Note that at very high frequencies the LL equation stops being applicable and our description is not valid anymore.}, two branches of AFM are converging and SW wavelengths $\lambda_{AFM\pm}$ are decreasing, as shown in Fig.~\ref{fig:2_a_b(inset)}(c). In Fig.~\ref{fig:2_a_b(inset)}(d) the dependence of $T^2$ on frequency $f$ is shown without multiplication by $\xi$ (for $\alpha_{ex,2}=0.3\alpha_{ex,1}$). This leads to $T^2 =0$ at the activation frequency $f_a$ for the plus AFM branch, and $T^2 =1$ for the minus AFM branch. The dependence $\xi (f)$ for both AFM branches is shown in the inset of Fig.~\ref{fig:2_a_b(inset)}(d). 

\begin{figure}[tbp]
	\centering
	\includegraphics[width = 0.95\linewidth]{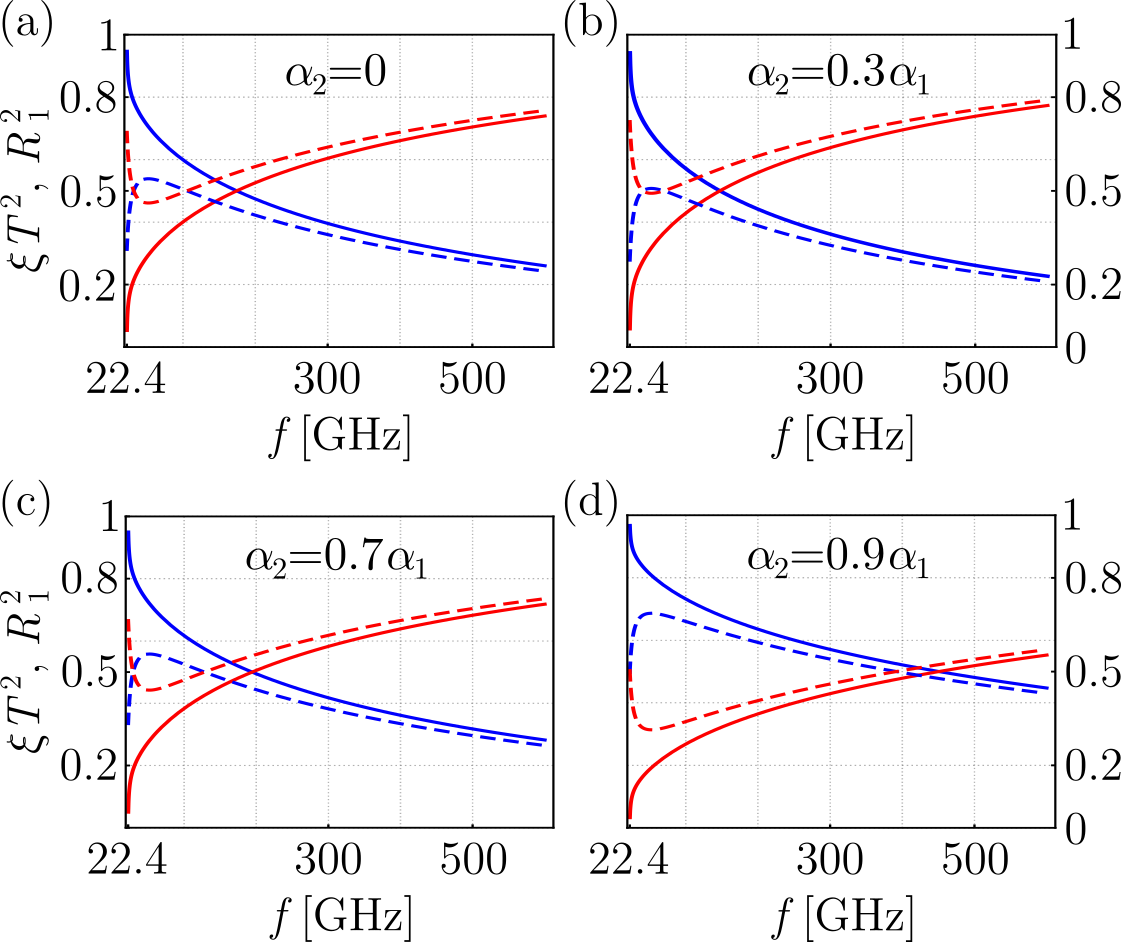}
\caption{Normalized transmittance $\xi T^2$ (red solid and dashed lines indicate plus and minus AFM branches, respectively) and reflectance $R_{1}^{2}$ (blue solid and dashed lines indicate plus and minus AFM branches, respectively) as functions of frequency $f$ for four cases: (a) the exchange energy between AFM sublattices at the interface is neglected $\alpha_2=0$, (b) $\alpha_2=0.3\alpha_1$, (c) $\alpha_2=0.7\alpha_1$, and (d) $\alpha_2=0.9\alpha_1$.}
\label{fig:fig4}
\end{figure}

To control SW propagation via material parameters tuning, it is convenient to consider the dependence of the physical transmittance $\xi T^2$ on frequency. We consider the vicinity of noncompensated case, i.e. when $\mu\ll 1$ \footnote{Despite the fact that the magnitudes of the magnetization in FM and sublattice magnetization in the AFM have to be approximately equal, i.e. $\nu \approx 1$, we have tested the limits of our model applicability for various values of $\nu$ within the range $[0.5,1.5 ]$, thus confirming that it only leads to quantitative variations.}. Since for $\alpha_{2}>\alpha_{1}$, the SWs in AFM cannot propagate, we only consider the range $\alpha_{2}=[0,\alpha_{1}]$.  The case $\alpha_{2}=0$ is shown in Fig.~\ref{fig:fig4}(a), where on the plus AFM branch the physical transmittance increases monotonously, while on the minus AFM branch it decreases up to a certain frequency and then increases again. Moreover, at high frequencies the physical transmittances for both AFM branches converge. At very higher frequencies ($\omega\sim$~6~THz), $\xi T^2$ reaches maximum value of 1 and then decreases to zero (and respectively the reflectance reaches minimum value of 0 and then increases at $\omega \to \infty$ to 1) \footnote{See Sec.~V of the Supplemental Material \cite{SupplMat}}. Therefore, the full SW reflection or full transmission can be achieved at certain $\omega$. The cases for different $\alpha_{2}/\alpha_{1}$ ratios are shown in Fig.~\ref{fig:fig4}(b)-(d). When $\alpha_{2}=0.3\alpha_{1}$, see Fig.~\ref{fig:fig4}(b), for the minus AFM branch, the physical transmittance does not drop below $0.5$. Note that at higher frequencies in this range, the physical transmittance is always larger than the reflectance at any $\alpha_{2}$. Meanwhile, to obtain $\xi T^2 > R_1^2$ in a lower-frequency range, it is enough to use the minus AFM branch, as shown in Fig.~\ref{fig:fig4}. The phase shifts between the incident and reflected SWs $\varphi_{1}$ and $\varphi_{2}$ are always equal to $\pi$, whereas the phase shift between the incident and transmitted SW $\varphi =0$.

\textit{Discussion.} Most of the aspects of our SW propagation theory are applicable for any values of the DSNA $\mu$, except for the boundary conditions (\ref{eq:eq2}), which are valid only in the vicinity of $\mu=0$, i.e., for nearly noncompensated AFM/FM interface. Therefore, if one finds the way to generalize these conditions to any $\mu$, the entire theory is immediately extended to any DSNA, thus covering the full range of AFM/FM interfaces: compensated, noncompensated, and anywhere in between. Then our model would also describe AFM/FM interfaces with not only flat profiles, but complex interfaces that take into account surface roughness, diagonal-like interfaces, and even curvilinear interfaces with variable DSNA. Furthermore, an analogous technique may be used to treat the SW propagation through AFM interfaces with other magnetic media.   

\textit{Conclusions.}  We have demonstrated that introducing a new physical characteristic of a finite-thickness interface, the DSNA $\mu$, makes an essential step towards describing a SW propagation through AFM/FM boundary. We have shown that in noncompensated case, by varying the exchange interaction parameter between AFM sublattices, $\alpha_{2}$, fine control of the SW transmittance through the AFM/FM interface can be achieved in the entire range from full reflection to full transmission depending on the SW frequency. This approach may open doors for tailoring interfaces with given AFM exchange constants for future magnonic nanodevices.

\textit{Acknowledgments.} 
The authors are grateful to Prof. Yu. Gorobets for very fruitful discussions, who passed away during the preparation of this paper. O.B. and O.G. acknowledge him as a mentor, his guidance and support continue to live on through everyone who knew him. O.B. acknowledges the support of this work by the Academy of Finland (Grants No.~320086 and No.~346035). O.A.T. acknowledges the support by the Australian Research Council (Grant No.~DP200101027), the Cooperative Research Project Program at the Research Institute of Electrical Communication, Tohoku University (Japan), and NCMAS grant.

\bibliography{spinwave_AFM}
\end{document}